# The Ouroboros Model

## Knud Thomsen

Paul Scherrer Insitut
5232 Villigen PSI
Switzerland
knud.thomsen@psi.ch

**Abstract**

At the core of the Ouroboros Model lies a self-referential recursive process with alternating phases of data acquisition and evaluation. Memory entries are organized in schemata. Activation at a time of part of a schema biases the whole structure and, in particular, missing features, thus triggering expectations. An iterative recursive monitor process termed 'consumption analysis' is then checking how well such expectations fit with successive activations. A measure for the goodness of fit, "emotion", provides feedback as (self-) monitoring signal. Contradictions between anticipations based on previous experience and actual current data are highlighted as well as minor gaps and deficits. The basic algorithm can be applied to goal directed movements as well as to abstract rational reasoning when weighing evidence for and against some remote theories. A sketch is provided how the Ouroboros Model can shed light on rather different characteristics of human behavior including learning and meta-learning. Partial implementations proved effective in dedicated safety systems.

## Introduction

The Ouroboros Model describes an algorithmic architecture for cognitive agents. Its venture point are two simple observations: animals and human beings are embodied, strongly interacting with their environment, and, they can only survive if they maintain a minimum of consistency in their behavior. The first issue poses severe constraints but also offers indispensable foundations for a bootstrapping mechanism leading from simple to sophisticated behaviors. As for bodily movement, also for cognition some measure of coherence and consistency is required, e.g. nobody can move a limb up and down simultaneously, and, at least in real-world settings, opposites cannot both be fully true at the same time.

Below, a brief outline of the algorithm will be presented followed by selected effects ascribed to the proposed structures. The emphasis here cannot lie on completeness but rather it is attempted to sketch an overarching picture covering and linking together a wide range of topics in a principled approach, suggesting one credible pattern.

## Action - and Memory Structure

Following Hebb's law, already in quite simple animals neurons concurrently active often experience an enhancement of their link raising the probability for later joint activation (Antonov, Antonova, Kandel, and Hawkins 2003). Different forms of learning act in concert. Neural assemblies are permanently linked together when once co-activated in the right manner. Later partial activation leads to an increase in the probability that the whole associated neural population fires together. Structured memories are laid down, and this especially effectively, when they are associated with some markers for success. According to the Ouroboros Model thus representations are bound together and preserved for later use, i.e. schemata joining diverse slots into cohesive memory structures are established. In rather direct extension of conditioned reflexes, the processes and the resulting memory entries are the same in principle, irrespective whether components of movements are combined into more elaborate choreographies, percepts into figures, or activations in many different brain areas into an entry for an episode. Data in brains are consequently organized into hierarchies of schemata where activation of any part promotes graded activation also for the rest of the linked features.

As a consequence of these structures, every neural activation triggers an expectation for the other constituents, which are usually active in this context. Activation at a time of part of a schema biases the whole structure with all associated slots and, in particular, missing features. Schemata can thus be seen as generalization of production rules (Anderson 1983).

It is important to note that in addition to old structures established well in advance due to the repeated connection of the involved attributes, schemata can also be generated on the fly, assembled from parts of existing building blocks as the occasion arises (Schacter and Addis 2007). For example, just mentioning in a conversation a "tall blond lady in a coat", immediately triggers a picture, which includes additional potential specifics, like that the lady wears some type of shoes, probably high heels or boots.

## Principal Algorithmic Structure

At the core of the Ouroboros Model lies a self-referential recursive process with alternating phases of data acquisition and evaluation. An iterative monitor process termed 'consumption analysis' is checking how well expectations triggered a one point in time fit with successive activations. A principal activity cycle is identified. Freezing for once the perpetually cycling activity and selecting an almost arbitrary starting point the following succession of steps can be outlined:

>... anticipation,
>action / perception,
>evaluation,
>anticipation,...

These sub-processes are linked into a full circle, and the snake bites its end, the Ouroboros devours its tail.

## Consumption Analysis

Any occurring activation excites associated schemata. The one with the highest activation is selected first. Taking the anticipations active at that time as reference and basis, consumption analysis checks how successive activations fit into this activated schema, i.e. how well low level input data are "consumed" by the selected frame structure. Attributes are 'explained away' (Yuille and Kersten 2006).

If everything fits perfectly the process comes to a momentary partly standstill and continues with new input data. If discrepancies surface they have an even more immediate impact on the following elicited actions. The appropriateness of schemata can vary in a wide range. In any case, consumption analysis delivers a gradual measure for the goodness of fit between expectations and actual inputs, in sum, the acceptability of an interpretation. Thresholds for this signal are set in terms of approval levels depending on the context; in the real world nothing can always be perfect, nevertheless, a wrong schema has to be discarded at some point.

Consumption analysis can be understood as a particular algorithm for pattern matching, which not only delivers simple feedback on identity but also some meta-information concerning overall performance, the quality of results, and suitable next actions in an actual situation.

## Selected Effects in the Ouroboros Model

Consumption analysis highlights discrepancies between anticipations based on prior experience and the current situation. As will be argued below, the very same basic processes works with goal directed movements, with conceptual plans and goals as well as with abstract rational reasoning when weighing evidence for and against some remote theories. A few prominent effects of the working of the proposed processes are sketched in the following.

## Attention

As an immediate consequence, occurring contradictions signal that something unexpected is encountered and that, against the background of experience, some modification of the current behavior might be necessary.

At the simplest level, a missing feature can be seen as effectively triggering attention to that open slot in a schema structure. Faces for example are looked at not in random scans but rather along paths linking most important features like eyes, mouth, ears and nose (Noton and Stark 1971). One feature directs attention to the next. Distinct discrepancies focus the gaze and mind on issues requiring further action. Unexpected peripherally perceived motion triggers attention and a saccade. Not only in natural scenes top-down guidance can overrun bottom-up saliency depending on the task (Einhäuser, Rutishauser, and Koch 2008). The amount of the resulting arousal can serve as a measure of assigned relevance; it depends in a meaningful manner on the history and the memory of the agent. If dimensions are involved which are considered important, minor deviations from expectations cause a big effect; discrepancies involving issues, which were unimportant or boring in the past, stir less attention. The weight of a mismatch is also self-reflectively modulated based on the actual situation including the associated urgency. If some parts of an activated schema fit very tightly while at the same time there are striking discrepancies involving other expected features this triggers more focused attention and causes a higher tension than if a schema fits overall with some evenly distributed minor deviations between anticipations and current observations. If the mentioned tall blond lady mentioned before turns out to actually have red hair this would attract some attention, if she had naked feet, probably even more.

Stumbling over an obstacle which usually does not lie in the middle of the way certainly draws the attention of the actor towards his feet and to the ground. The effect is basically the same when one hears a politician making a statement running counter to expectations derived from knowing the established consensus in a party.

## Emotion

Monitoring the quality of congruence with experience by the consumption analysis provides a very useful feedback signal for any actor under all circumstances. Assuming on average something like 80 percent agreement, this would set the zero level of the scale: deviations to the worse as well as to the better are worth while highlighting and these episodes should preferentially be remembered.

It is claimed that emotions are primarily that: feedback signals from a consumption analysis process to the actor, he is deeply involved.

Emotions occur as a consequence of a situation or an event and they set the stage for activities following thereafter. This picture combines well established and seemingly contradictory stances explaining emotions (Scherer 1999; Ekman 1999). Affects as information resulting from appraisal the same as motivational accounts stressing behavioral dispositions are just different facets of the basic process in the Ouroboros Model. When discrepancies become too big, an interruption, a kind of reset, of the ongoing activity will be triggered. Emotions appear to be tied to a particular event, object or situation; moods usually denote less focused affects and circumstances. In addition, these manifestations of one and the same basic signal differ in their time characteristics.

Consistency in the Ouroboros Model is checked globally, in parallel for all features, but as discrepancies are dealt with according to their weight, a rather fixed sequence of appraisal dimensions might be observed because of general similarities between schemata, i.e. shared parts (Sander, Grandjean, and Scherer 2005).

Actors are often not alone, any signal useful for one, certainly is of relevance also to others. This would explain the communication value of displayed emotions.

Similarly as observed for the weight of dimensions for attracting attention, also emotions come in two versions (and their combinations). They can be inherited, i.e. forming a constitutive part of a schema as an earlier associated feature; in this case they can be activated very directly and quickly. When novel circumstances give rise to a never before experienced evaluation, attention will be evoked and corresponding emotions will build up; they are incorporated into the memory of the event, – ready for later fast use.

**Problem Solving**

In many circumstances a Bayesian approach can be verified as the optimum way of considering all available evidence in order to arrive at a decision, e.g. in classification and learning (Tenenbaum, Griffiths, and Kemp 2006; Yuille and Kersten 2006). In all cases it is most important to combine prior probabilities with current data. The interplay between a partially activated schemata and newly observed features does just that. When certain combinations in general are much more likely than particular others, less additional evidence is required in the former case than in the latter.

Knowledge guided perception and the organization of knowledge in useful chunks, i.e. meaningful schemata, was found to lie at the basis of expert performance (Ross 2006). Beyond the effectiveness of the proposed memory structures in the Ouroboros Model for the selection of interpretations, the whole process of directed progress of activation in a mind is well adapted to more abstract problem solving. Constraint satisfaction has been proposed as a general process for rational behavior and reasoning, applicable in a very wide range of settings (Thagard 2000); maximizing the satisfaction of a set of constraints can be seen as maximizing coherence (Thagard and Verbeurgt 1998). The Ouroboros Model and consumption analysis not only offer an efficient implementation of constraint satisfaction but also provide a rationale and a basis for goal directed refinement and self-steered tuning of the process.

The succession of parallel data acquisition and –evaluation phases interspersed with singular decision points in an overall serial repetitive process offers a natural explanation of purportedly contradictory observations concerning the prevalent character of human data processing: competing views stressing serial or parallel aspects can both be correct, depending on the investigated details.

Attention is quickly and pre-consciously focused on the most pressing questions at a time. Emotions provide the feedback how well things go; they are a constituting ingredient to rational behavior. Depending on the history and personal preferences of an agent, all of this is to some degree individual, more or less variable for different persons. In any case, a minimum of the slots deemed important have to be filled satisfactorily to accept a solution. Most extensive achievable consistency is the main criterion for judging the value and reliability of primary sensory percepts, and even more so, the "truth" of theories of highest complexity, abstraction and remoteness. As a corollary, the Ouroboros Model naturally also explains observations, which are taken by some as evidence that "non-classical", in particular proper quantum effects, would be required for consciousness (Manousakis 2007). Seemingly mimicking quantum behavior, in very general terms, after a phase of primarily parallel data collection the process settles on one interpretation: a "superposition" "collapses" non-linearly (Thomsen 2008).

**Cognitive Growth**

With a positive signal that everything fits nicely the associated positive emotions mark a good basis concerning the expected future usefulness of the schema in question.

Whenever pre-existing structures cannot satisfactorily accommodate new data this will be accompanied by another clear signal from the consumption analysis monitor. Negative emotions notify the need to change something, if possible; – at least the assessment of the input data, or new additional schema structures.

In both cases, effective self-regulated bootstrapping occurs. Useful schemata will be enhanced and better learned; they can serve as material for future refinement and as building blocks for more sophisticated concepts. Features where discrepancies surfaced are identified. Extended or completely new schemata will be the consequence (Piaget and Inhelder 1986). Learning thus does not take place indiscriminately but at the right spots where it is most effective. Cognitive growth preferably occurs exactly when and where it is needed. The Ouroboros Model comprises an efficient learning strategy; in fact, inherent meta-cognition allocates resources and directs the effort.

**Teaching**

New concepts and content can best be absorbed if they lie just at the boundary of the already established structures; only there, gaps are rather well defined and success is in easy reach when filling in those vacancies. This stresses the importance of self-paced exploration and of fun associated with the activity. Montessori education appears to emphasize these points; it is reported to yield impressive results (Lillard and Else-Quest 2006). Effective teaching thus builds on the processes outlined above. Natural and artificial minds can best be taught when pre-existing knowledge structures and novel input are carefully matched. A provoked aha-experience in all cases marks some content as useful, and it is an efficient trigger for preferred long term storage. Victory over a competitor with about equal strength might provide strong additional motivation (Ross 2006).

## Implications for Human Behavior

Although already above in the discussion of selected effects claimed to be explained by the Ouroboros Model, human behavior delivered examples for specific findings; here, the focus is further shifted towards the human. First, a few selected experienced shortcomings and imperfections of the behavior of human agents will be investigated, and it will be shown how the before described structures and processes can shed light on rather diverse characteristics, which by many are only ascribed to humans, probably animals, and principally not to machines.

### Priming / Masking

Triggering an expectation linked to a specific location facilitates acting on the highlighted information. This is one basic finding of simple priming experiments; another one concerns semantic priming where reactions are faster when an appropriate context is activated. These effects can naturally be ascribed to the biasing influences of schemata. After features are successfully consumed, they have to be marked in order to avoid that they are improperly considered again. Attentional blink occurs when subjects watch for targets in a stream of stimuli: they likely miss a target if it follows with a short delay after a first and detected target.

Distracting subjects with task-irrelevant information attenuates the attentional blink (Olivers and Nieuwenhuis 2005); as more than one activity is going on, consumption tagging is less efficient. In a straight forward manner the Ouroboros Model can deliver a unified account for competing models of attentional blink: resource depletion, processing bottleneck, and temporary loss of control can all be understood as different facets originating in one process (Kawahara, Enns, and Di Lollo 2006).

The proposed periodic processing entails a marked structure with respect to time, and its perception. The shortest time span that can be discerned by humans has been found to be approximately 30 ms (Pöppel 1997). It seems obvious, that this would be the elementary period of the loop of consumption analysis.

If simply the time available is not is not enough to conclude at the minimum one full circle because the sensory input is quickly replaced by a mask, no complete perception can be obtained. Difficult to perceive stimuli requiring more than one cycle would correspondingly be sensitive to disruption for a longer period.

Under special conditions negative congruency effects can be observed where subjects respond faster to incongruent prime stimuli compared to congruent ones. This happens when the prime is consuming already fitting parts of the mask and new priming is elicited from the mask's remainders, which are again similar to the target (Kiesel, Berner, and Kunde 2006; Tononi and Koch 2008).

Non-overlapping masks, which terminate at the same time as the target, can impede target perception even if their presentation started only after the target; this can be explained by a more general reset, when the inputs for many slots become concurrently unavailable.

### Sleep

All known agents who are able to exhibit some substantial measure of intelligence and consciousness spend a sizeable fraction of their life in strange states of sleep. Any attempt at a comprehensive account of mental functions should therefore include a profound explanation of this fact.

Given stringent time constraints in the real world concerned with survival, consumption analysis inevitably produces "leftovers", i.e. not-allocated features and not-confirmed concepts, which accumulate with time. The Ouroboros Model explains sleep as a specific and multifaceted housekeeping function for maintaining appropriate signal / noise conditions in the brain by actively erasing data garbage. Knowing that data clearing is a thermodynamically irreversible operation requiring energy (Bennett 1987), it does not surprise that general activation levels in the brain are comparably high during waking and sleeping.

Many at first sight rather diverse observed characteristics and proposed functions of sleeping and dreaming can be seen as consequences of efficient data processing and appropriate clean-up:

- extra and especially novel activities necessitate an increased need for sleep as the involved schemata are overflowing more, in particular when they are still developing;
- for the same reason babies and children need more sleep than adults;
- disturbing, threatening and unresolved issues would predominantly surface as dream content as their processing could not be concluded yet;
- episodes which stir emotions, because expectations – e.g. norms – are violated, will preferably provide material for dreams;
- erasing erroneous associations and traces of expectations or of perceptions unaccounted for enhances correct and well established connections; the ensuing increase in signal / noise would make memory entries, checked for consistency by the consumption analysis, stand out and thus easier to retrieve and reactivate;
- the variations in neuromodulator levels during sleep phases reported in the literature appear to be well in line with their proposed role in "cleaning up";
- the paradoxical activity of brain cells and areas during sleep seem to fit under this account with the functions which they are assumed to perform during awake behavior.

One mechanism of increasing available capacities and enhancing the accessibility of memories via improvements in signal / noise does of course not exclude that there are other, even more direct, processes also happening during sleep which can also boost wake performance.

**From Self Awareness to Consciousness**

We are embodied agents. Our body anchors and grounds each of us firmly in the real world; – not only statically but, most importantly, with every dynamic action. Self-awareness and consciousness come in shades of grey; they start with physiology, somatic proprioception and a sense of ownership for an action. Drawing on multiple and multimodal sources a sensory event is attributed to one's own agency when predicted and assessed states are congruent (Synofzik, Vosgerau, and Newen 2007). First intentions do not need much second-order thought: a goal, like for a simple reaching movements, is aimed at, and it is reached or corrections are necessary.

There is convincing experimental evidence that awareness of an action stems from a dynamic fusion of predictive and inferential processes (Moore and Haggard 2006). It has been suggested that pre-motor processes lay down predictive signals in anticipation of effects, which are integrated into a coherent conscious experience encompassing actions and effects as soon as feedback sensory input becomes available (Haggard and Cole 2007). Dissociations between focused attention and consciousness have been found under special conditions (Koch and Tsuchiya 2006). At the coarse level intended here three conditions can be distinguished in the light of the Ouroboros Model:
- Perfect fit between expectation and data;
- Deviations in the expected range;
- No acceptable match.

If the first (unconscious) guess on what frame a certain feature might belong to was right, the recursive process will quickly converge and result in one strongly activated concept. Consumption analysis then yields that all features are nicely consumed and nothing is left over; all data are consistent and coherently linked in the selected interpretation. Aha-experiences relevant to the actor catch attention and become conscious by evoking higher order personality activation "HOPA" connecting the context and details of the task and also of the actor. The break that follows a moment where no continued attention is required might just be what is needed to allow the current general activation to grow to reach a threshold for consciousness (Tononi and Koch 2008). As proved useful, these episodes are preferably committed to long term memory.

If in a specific situation topical "local" information does not suffice to obtain a unique and coherent interpretation, activation spreads and more remote content is considered. Data from the episodic memory will always contain representations pertaining to the body and thus some anchoring to the actor. Thus she is becoming more involved as the need for a solution becomes more pressing and activation soars. With high enough weight, significant gaps in the understanding of a situation will thus trigger HOPA, too.

In between the above two sketched extremes lies much "thought-less" action, e.g. driving home the usual way without any special event; – this could be called a Zombie-mode of operation.

Some higher order personality activation including representations and signals of the body of the actor is claimed to lie at the basis of consciousness. If possible, even more than in other global theories of consciousness (Baars 1988; Dehaene, Kersberg, and Changeux 1998; van Gulick 2004) the all-embracing involvement of the actor herself is considered most important, which is accomplished by wide broadcasting and requesting of input; this becomes especially evident in the cases where the purpose of the global excitement is to bring to bear all possibly useful information in a difficult situation.

The memory entries for episodes in turn contribute to the life story of the actor, his self-concept, the narrative self. References to the body provide privileged semantic content and anchoring; they ensure continuity and the uniqueness of a record (Baars 1988; Pauen 2007; Pöppel 1997).

Inasmuch as a direct and un-reflected impulse for action, provoked by some distinct trigger and fuelled by a strong associated emotion, is moderated and subject to a second consideration, where a wider context is taken into account, awareness and conscious elaboration can improve the performance of an actor (Lambie 2007; DeWall, Baumeister, and Masicampo 2008).

Almost the same is actually true also the other way round: existing schemata can produce an undue focus or blockage; self-aspects themselves also bind processing capacities, which can lead to sub-optimum choices following conscious deliberations as compared to "gut decisions" (Dijksterhuis, Bos, Nordgren, and van Baeren 2006). It is important to note that in the reported tasks demonstrating the superiority of "unconscious deliberation" all required information had been given before.

A special quality emerges as soon as information about the owner of the processes is self-reflexively included and associated representations, i.e. autobiographic memory and self models, are embraced in the focus of the same basic processing. When the actor himself is in the center of his attention, other content fades, loses importance and weight. This starts when looking at ones toes and continues to recollections and reflections on personal experiences, goals, plans, preferences and attitudes. Deeply anchored in our whole body, this self-reflective and self-relevant recursively looping activity in the brain is drawing on and again inducing a sense of ownership for (mental) actions as well as "qualia" and all associated emotions; – our total personal experience. Everybody who has ever burnt his finger will not primarily be bothered by doubts concerning the true essence or general validity of temperature sensations; he will certainly know what heat feels like.

Notwithstanding that it was found that not all shades of consciousness are depending on language, details subject to definitions, it certainly is the most powerful tool when it comes to consciously assembling and manipulating elaborate models in a mind, – not to speak of deliberations or their communication (Tononi and Koch 2008).

Emphasising diverse aspects, definite properties can be ascribed to the complex and multifaceted phenomenon of a self, which is seen by many authors as providing the unifying thread from episodic memory to consciousness and free will (Samsonovich and Nadel 2005).

**Pleasure in Play and Art**

Perception like memory retrieval or movement is an active process. One feature which is preferably included in relevant schemata relates to the effort usually required and the time needed to perform an action. Also the consumption of these meta-features is monitored, and these slots can be filled with little, as expected, or more than anticipated problems. This self-monitoring for (perceptual) fluency is seen as the basis for pleasure in play and art in the Ouroboros Model (Thomsen 2000; Reber, Schwarz, and Winkielman 2004).

## Real-World Applications

In three different dedicated safety systems very limited aspects of the Ouroboros Model have already successfully been implemented. In principle, these systems featured a main loop for data acquisition, their evaluation with respect to some criteria, and actions depending on the results including the provision of intermittently updated signals about the health status of the systems and the actual quality of their performance, used for on-line adaptation of their behavior.

**LASSY.** A Laser Surveillance System for spent nuclear fuel distilled most simple categories from measurements, which were used in an adaptive manner. If one cycle gave inconclusive results, "attention" and a second look were triggered (Thomsen, Morandi, Sorel, and Hammer 1989).

**VIMOS.** A Visual Monitoring System was built to prevent a severe accident in the SINQ spallation neutron source at the Paul Scherrer Institute (PSI). The sensitivity of the system derives partially from making use of carefully crafted expectation values, i.e. the specific patterns signifying dangerous constellations are well known in advance, and this knowledge together with accruing experience is used to continually optimize interlock criterions during operation (Thomsen 2007).

**Leak Detection.** Reliable leak detection was another mandatory requirement when operating an especially delicate liquid metal target in the SINQ facility (Thomsen 2008). When significant and consistent deviations from the original expectations were observed, new optimized and dynamically established reference values for measured temperatures secured the full functionality of the device.

Although in all three cases the meta-loop for category construction was not implemented in software but structures were provided by the user, there is no doubt that full automation would have been easily feasible.

## Discussion

It goes without saying that the here presented sketch of the Ouroboros Model outlining its basic working and some suggested consequences cannot provide more than just a glimpse of a comprehensive theory, which will need much more detail and quantitative elaboration. The very object of this short paper is to draft an overview of what a large puzzle could look like. Working out the intricate details should follow and be guided by expectations, the gist, briefly indicated here.

Of diverse anticipated objections to the Ouroboros Model only one shall be addressed shortly. It is most important to point out that there arises no problem due to seemingly circular arguments. The principal recursive algorithm progresses and evolves *in time*. When the snake bites its tail, the teeth and the tip of the tail belong to two distinctly different points in time. Starting with any basic set of expectations, discrepancies to any actual input data can be determined and used for choosing and steering the following actions including the establishment of revised expectations. This full cycle can in fact not be depicted meaningfully by a circle but more faithfully by a spiral. The intrinsically built-in potential for growth out of whatever plain of current understanding to novel heights, opening new dimensions in a productive way is one of the key and most valuable features of the Ouroboros Model.

## Future Directions

Some of the described characteristics, like the working of a principal basic consumption analysis loop, appear to the author as unavoidable for efficient actors in the real world, while others, like the need for sleep could be just implementation detail. Looking at the sheer size of the brains of orcas, which can do for long periods without sleep, one could imagine that diluting data garbage and waiting for slow decay might be an alternative to more active disposal and erasure.

Going a very long way beyond the first primitive implementations in existing safety systems, sophisticated

hard- and software agents would raise questions related to free will and morals (Anderson and Leigh 2007). It can be agued that the observation of contradictions – to established general principals or to patterns derived from selected examples, while paying heed to anticipated consequences and explicitly taking decisions on this basis – would be firm ground and a suitable venture point for understanding natural as well as building artificial true intelligent and conscious agents.